\documentclass[twocolumn,twoside,slac_two]{revtex4}

\usepackage{graphicx}
\usepackage{fancyhdr}
\usepackage{xspace}

\newcommand{\passsix}{Pass~6\xspace}
\newcommand{\passsixhyphen}{Pass-6\xspace}
\newcommand{\passseven}{Pass~7\xspace}
\newcommand{\passsevenhyphen}{Pass-7\xspace}
\newcommand{\passeight}{Pass~8\xspace}
\newcommand{\passeighthyphen}{Pass-8\xspace}

\newcommand{\Fermi}{{\it Fermi}\xspace}

\begin{document}

\title{Pass 8: Toward the Full Realization of the \Fermi-LAT Scientific
  Potential}

\author{W.~Atwood}
\affiliation{Santa Cruz Institute for Particle Physics, University of
  California, Santa Cruz}
\author{A.~Albert}
\affiliation{The Ohio State University}
\author{L.~Baldini}
\author{M.~Tinivella}
\affiliation{Universit\`a di Pisa and INFN-Sezione di Pisa}
\author{J.~Bregeon}
\author{M.~Pesce-Rollins}
\author{C.~Sgr\`o}
\affiliation{INFN-Sezione di Pisa}
\author{P.~Bruel}
\affiliation{{}LLR, Ecole polytechnique, IN2P3/CNRS}
\author{E.~Charles}
\author{A.~Drlica-Wagner}
\author{A.~Franckowiak}
\author{T.~Jogler}
\author{L.~Rochester}
\author{T.~Usher}
\author{M.~Wood}
\affiliation{SLAC National Accelerator Laboratory}
\author{J.~Cohen-Tanugi}
\affiliation{Laboratoire Univers et Particules de Montpellier,
  Universit\'e Montpellier 2, CNRS/IN2P3}
\author{S.~Zimmer}
\affiliation{The Oskar Klein Centre for Cosmoparticle Physics}
\author{for the \Fermi-LAT Collaboration}

\begin{abstract}
The event selection developed for the \Fermi Large Area Telescope before
launch has been periodically updated to reflect the constantly improving
knowledge of the detector and the environment in which it operates. \passseven,
released to the public in August 2011, represents the most recent major
iteration of this incremental process.

In parallel, the LAT team has undertaken a coherent long-term effort aimed
at a radical revision of the entire event-level analysis, based on the
experience gained in the prime phase of the mission. This includes virtually
every aspect of the data reduction process, from the simulation of the
detector to the event reconstruction and the background rejection. The
potential improvements include (but are not limited to) a significant
reduction in background contamination coupled with an increased effective
area, a better point-spread function, a better understanding of the systematic
uncertainties and an extension of the energy reach for the photon analysis
below 100~MeV and above a few hundred GeV.

We present an overview of the work that has been done or is ongoing and the
prospects for the near future.
\end{abstract}

\maketitle

\thispagestyle{fancy}

\section{Introduction\label{sec:intro}}

The current%
\footnote{Hereafter we use the word \emph{current} to refer to the specific
  event-level analysis and associated event classes publicly available at the
  time of writing, i.e. \passseven version~6, detailed
  in~\citet{REF.2012.P7PerfPaper}.}
Large Area Telescope (LAT) event-level analysis framework was largely developed
before launch using Monte Carlo simulations through a series of iterations
that, for historical reasons, we call \emph{Passes}.
\passsix was released at launch, and was followed in August 2011 by \passseven,
which mitigates the impact of some of the limitations of its predecessor.
We refer the reader to \cite{REF.2009.LATPaper} for a description of the LAT
instrument and to \cite{REF.2012.P7PerfPaper} for detailed information about
the associated Instrument Response Functions (IRFs).

Though the current simulation and reconstruction framework has served adequately
the science of the prime phase of the mission, on-orbit experience with the
fully integrated detector has revealed some neglected and overlooked
issues---primarily the effect of the instrumental pile-up (hereafter
\emph{ghost events} or \emph{ghost signals}, see \citet{REF.2012.P7PerfPaper}
for more details). Not long after launch, opportunities for clear improvements,
with the potential to greatly extend the LAT science capabilities, were
identified in the three main areas of the event-level analysis: (i) Monte Carlo
simulation of the detector, (ii) event reconstruction and (iii) background
rejection~\citep{REF.2010.Pass8}.
These improvements are now being deployed in a systematic and coherent
fashion in the context of the \passeighthyphen iteration of the LAT event-level
analysis.

\section{Event Reconstruction\label{sec:recon}}

The event reconstruction is so far the single development area where most of the
effort has been put since the \passeighthyphen project started at the end
of 2009. In this section we review the main improvements we have implemented at
the reconstruction level for each of the three LAT subsystems: tracker,
calorimeter and Anti-Coincidence Detector (ACD).

\subsection{Tracker Reconstruction\label{subsec:recon_tkr}}

The conceptual model underlying the current reconstruction in the LAT tracker
is that a photon converts to produce an electron-positron pair and that the
electron and positron make hits in the tracker, which can be followed from the
point of conversion to their exit from the tracker. The present tracker
reconstruction code uses a track-by-track \emph{combinatoric}
pattern-recognition algorithm to find and fit the two tracks representing the
electron-positron pair, and then combines them to form a vertex representing
the photon conversion point.

This approach is problematic in four general areas.
First, the track-following algorithm needs an initial direction to start the
track following and hit attachment. To reduce the number of potential hit
combinations it uses the reconstructed calorimeter energy centroid and axis of
energy flow to choose the initial hits, and this makes the efficiency of the
track finding dependent on the accuracy of the calorimeter reconstruction
(as we shall see in Section~\ref{subsec:recon_cal}, the presence of ghost
signals in the calorimeter exacerbates the situation).
Second, the track model includes multiple coulomb scattering, which requires 
an estimate of the track energy, also derived from the calorimeter. 
Third, electrons and positrons interact readily in the material comprising the
tracker (in particular the converter foils and silicon), with the result that
photon conversions rarely resemble clean two-track events; instead they
produce multiple hits as the electromagnetic shower begins to develop.
Finally, large energy deposits in the calorimeter, particularly from
high-energy photons far off-axis, can result in \emph{backsplash}---particles
moving upwards from the calorimeter causing a large number of randomly hit
strips in the lower planes of the tracker. As the number
of backsplash hits increases the current track finding can become confused,
particularly if the initial position and direction estimates from the
calorimeter are not accurate. For photons that convert in the back (lower part)
of the tracker, where the real gamma-ray track can feature as few as three hits
in each orthogonal view, the potential to drown out real signal in a sea of
spurious hits is large.

These features combine to produce three main effects: the loss of the events
that fail to reconstruct at all, the migration of events from the core of the
Point-Spread Function (PSF) to the tails because of poorly reconstructed tracks
and increased confusion often resulting in mislabeling good gamma ray events
as background.

\subsubsection{Tree-based Tracking}

The \passeighthyphen reconstruction addresses these issues by introducing a
\emph{global} approach, called tree-based tracking, which looks at the
conversion in the tracker as the start of a shower and attempts to model this
process by linking hits together into one or more tree-like structures.
For each tree, the primary and secondary branches, defined as the two longest
and straightest, represent the primary electron and positron trajectories
(if unique) and sub-branches represent associated hits as the electron and
positron radiate energy traversing the tracker. Once the tree has been
constructed, its axis can be found by calculating the moments of inertia of
the associated hits, with the \emph{mass} of each hit taken as a function of
the length and straightness of its branch.
The axis can then be used to associate the tree to a particular cluster in the
calorimeter, which allows an estimate of the energy associated with the tree.
Once an energy is available, up to two tracks can be extracted, and fit, from
the tree by associating the hits along the primary and secondary branches.
As the tracks are extracted, they are fit using a Kalman Filter technique which
accounts for multiple scattering but also allows for kinks which exceed
deviations expected from multiple scattering and with each track given half
the energy from the associated cluster. The fit tracks incorporate the best
information about the axis of the tree, and provide the best resolution on the
direction of the photon.

Tests with Monte Carlo simulations and flight data show that the new
tracker pattern recognition has the potential to significantly reduce the
fraction of mis-tracked event and provide a 15--20\% increase of the
high-energy acceptance, with even larger improvement in the off-axis effective
area, especially for photons converting in the lower part of the tracker.

\subsubsection{Vertexing}

If more than one track is produced from a given tree, then an attempt is made
to combine the two tracks into the vertex expected in a pair conversion.
The track association in the vertexing stage exploits both the distance of
closest approach of the projected tracks (typically 1~mm or less for
genuine pair conversions) and the number of layers that separate the heads of
the tracks (either 0 or 1). The track fitting yields not only track parameters,
but also the associated errors (see Section~\ref{subsec:event_by_event_errors}
for more details). Accordingly, when the tracks are combined to form the
vertex, these errors are used to properly weight the contributions of each
track. The resulting vertex generally yields the best information on the photon
direction.

Since the multiple Coulomb scattering scales as $E^{-1}$, the weights input to
the vertexing (in the scattering-dominated regime) are proportional to the
square of the assigned track energies. Unfortunately the LAT has very limited
ability to determine the individual track energies, since the energy deposits
in the calorimeter associated with the individual tracks usually overlap.
In the context of \passeight the strategy to determine the energy ordering of
the tracks has been completely reworked and we anticipate that the improvement
achieved over the current ordering scheme will be reflected in a better overall
PSF.

\subsection{Calorimeter Reconstruction\label{subsec:recon_cal}}

Just as for the tracker, the calorimeter reconstruction has been extensively
revisited at all levels---from the crystal simulation (where, e.g.,
unaccounted-for end effects in the light collection proved to have noticeable
consequences on the high-level science performance) to the energy and direction
reconstruction and the background-rejection algorithms.

\subsubsection{Clustering}

The single most radical change is the abandonment of the single-particle
paradigm common to the pre-\passeighthyphen event-level analysis and the
introduction of a clustering stage in the calorimeter.
In the \passsixhyphen and \passsevenhyphen event reconstruction we treat the
energy deposit in the calorimeter as a monolithic entity, with all the hit
crystals grouped together. 
In a low occupancy environment such as the one in which \Fermi operates, this
approach proved to be adequate to support the science analysis of the prime
phase of the mission. However residual ghost signals away from the gamma-ray
shower (and therefore easily distinguishable from it, in principle) can
introduce substantial errors in the measurement of the energy, centroid and
direction of the shower itself. Since the matching between the tracker and the
calorimeter is one of the main inputs to the background rejection, this results
in a net loss of effective area from genuine gamma-ray events misclassified as
background%
\footnote{This loss was quantified and accounted for (but not recovered) in the
  generation of the post-launch IRFs, both in the \passsixhyphen and
  \passsevenhyphen flavors.}.

In \passeight we have introduced a clustering stage in the calorimeter aimed at
identifying the ghost signals and recovering the aforementioned loss in
effective area. This presents some unique challenges, mostly connected with the
fact that the LAT is designed to trigger on events over a huge field of view
and therefore the calorimeter is seldom projective. We therefore decided to take
advantage of the intrinsically three-dimensional calorimeter readout and
exploit a Minimum Spanning Tree (MST) construction---a concept borrowed from
graph theory with a long standing connection with clustering applications.

Tests performed on Monte Carlo simulations and flight data clearly confirm
the effectiveness of this approach, indicating a 5--10\% increase in the
effective area above $\sim 1$~GeV and, potentially, a much larger effect below
a few hundred~MeV, where the energy in the ghost signal can be of the same order
of magnitude or larger than that of the triggering gamma-ray.

\subsubsection{Energy Reconstruction}

The other crucial calorimeter-related development area is the energy
reconstruction at very high energy. Above a few GeV our workhorse
reconstruction method is a three-dimensional profile fit to the calorimeter
layer energies. This approach proved to be nearly optimal up to ~1 TeV, where
the average energy-release per crystal at the shower maximum starts exceeding
the dynamic range of the readout electronics and saturation becomes an issue.

In order to overcome this limitation, the profile fit has been extensively
reworked by breaking up the layer contributions into individual crystal
energies. Monte Carlo simulations indicate that the energy deposits in the
saturated channels can be recovered to some extent by using the information
from the nearby (non-saturated) crystals, achieving a decent energy resolution
up to $\sim 3$~TeV~\citep{REF.2012.EnergyRecon}.

\subsection{ACD Reconstruction}

The ACD reconstruction has been fully re-written in the context of \passeight.
The first major improvement comes from the novel incorporation of calorimeter
information when associating incident particle direction with energy deposition
in the ACD. Directional information derived from calorimeter clusters is now
propagated to the ACD in addition to tracks derived from the tracker. This
additional calorimeter information is particularly important for identifying
background events at high energies or large incident angles, which are more
susceptible to tracking errors. In these cases, the calorimeter provides the
more robust directional information. 

The second major improvement is to utilize event-by-event directional 
uncertainties when associating tracks and clusters with energy depositions in 
the ACD. Previously, the ACD reconstruction scaled the tracking uncertainty in
an \textit{ad hoc} manner based on an estimate of the total event energy.
However, widely varying event topologies can lead to large differences in the
quality of directional reconstruction for events of the same energy (see
Section~\ref{subsec:event_by_event_errors}). Capturing this information in the
event-by-event uncertainties provides substantially more information for
background rejection.

The third major improvement comes from utilizing the fast ACD signals provided 
to the LAT hardware trigger to mitigate the impact of ghost signals in the
slower ACD pulse-height measurements. This improvement is especially important
at low energies, where calorimeter backsplash is minimal and a small deposition
of energy in the ACD can lead to the rejection of an event. In this regime the
use of trigger information in the background rejection removes out-of-time
signals from the ACD and provides a significant increases in effective area.

\section{Event selection\label{sec:event_selection}}

As the event reconstruction is nearing completion, the LAT Collaboration
is now focusing on the next step of the event-level analysis, namely that
responsible for the final determination of the high-level event properties,
such as particle type, energy and direction.

Similarly to what we did for the previous passes, we use Classification Trees
(CTs) to select candidate gamma-rays on the basis of the reconstruction
outputs. The particle identification CTs are trained using variables from all
the three LAT subsystems.
One noticeable difference is the use of the TMVA multivariate analysis
framework \citep{REF.2007.TMVA}. Compared to the technology used in the current
event classification, TMVA is capable of handling much larger data sets and
allows for an overall faster development cycle.
The CT performance is evaluated from the combination of background rate and
gamma-ray acceptance that can be achieved for a given cut on the output signal
probability. We note here that a differential background rate equal or slightly
lower than the Extragalactic Gamma-ray Background (EGB) rate is desirable for
point-source analysis.

\begin{figure}
  \centering\includegraphics[width=0.93\linewidth]{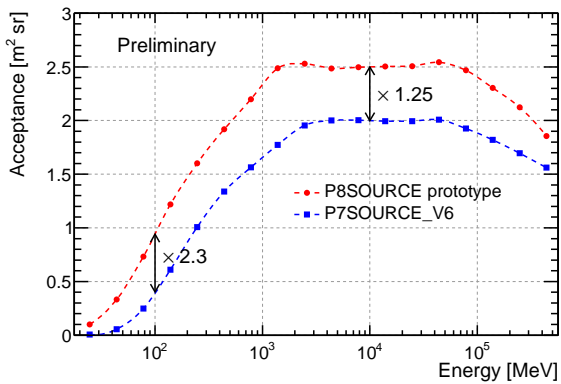}

  \caption{Gamma-ray acceptance versus photon energy for the \passsevenhyphen
    source class and a candidate \passeighthyphen event class.}
  \label{fig:p8_acceptance}
\end{figure}

We have studied several candidate \passeighthyphen event classes defined by
event selections that allow varying levels of background contamination relative
to the EGB. Each event class is composed of the following cuts: a selection on
events with a reconstructed track that deposits at least 5~MeV in the
calorimeter, an ACD pre-selection on events for which the reconstructed track
points to an activated section of the ACD, and an energy-dependent cut on CT
variables for the particle type and the quality of the angular
reconstruction. As shown in Figure~\ref{fig:p8_acceptance}, at high energy we
find a $\sim 25\%$ increase in acceptance relative to the \passsevenhyphen
source event class while at low energies (below $\sim 300$~MeV) the increase in
acceptance can be as high as a factor three.

\section{Extended Event Classes}
 
In the current photon analysis, events with no track in the tracker or
depositing less than 5~MeV in the calorimeter are simply discarded---though
some of them are used in non-standard analyses such as the LAT Low-Energy
technique (LLE) described in \citet{REF.2010.LLE}.
One of the areas of improvement which is being investigated in the context of
\passeight is the development of extended photon classes, e.g.
tracker-only and calorimeter-only events. Although these events have worse
energy resolution and/or PSF with respect to those in the standard photon
classes, they can provide a very significant increase of effective area
in some regions of the LAT phase space that might be exploited in specific
science analyses.

\subsection{Tracker-only Events}

Below $\sim 100$~MeV many of the $e^+$ and $e^-$ range out in the tracker and
deposit no energy in the calorimeter. Since seeing calorimeter signals
that are correlated with tracks is a powerful discriminator for
well-reconstructed events and helps in rejecting particle backgrounds that 
would otherwise be difficult to identify, in the current event selection we
do require a minimum (5~MeV) energy deposit in the calorimeter.
However, the success of the LLE analysis for bright transients has made it
clear that tracker-only events carry useful information that can be used in
science analysis.

As the entire event selection process is now being re-assessed in the context
of \passeight, tracker-only events provide the potential for a substantial
increase in the effective area below 100~MeV, opening a region of the
LAT phase space which is extremely interesting for many science analyses.

\subsection{Calorimeter-only Events}

While almost a half of the events above $\sim 50$~GeV%
\footnote{Technically, a significant fraction of these events would be
  discarded by the onboard filter, if this was not disengaged for events
  \emph{depositing} more than 20~GeV in the calorimeter. At these energies the
  shower leakage is such that the 20~GeV (of deposited energy) high-pass
  threshold translates into a smooth effective threshold of $\sim 50$~GeV,
  when measured in reconstructed energy.}
have no usable tracker information (either because they convert in the
calorimeter or due to mis-tracking), at these energies the LAT calorimeter
provides a directional capability at the level of a few degrees or better.

Although this is much less precise than the typical tracker PSF,
calorimeter-only events constitute a very promising event class for those
analyses where the pointing accuracy is not critical. Preliminary simulations
show that they might provide as much as $\sim 30\%$ increase in the high-energy
acceptance, with an even larger ehnancement of the effective area at large
off-axis angles. The rejection of particle backgrounds in the absence of usable
tracker information must still be studied in detail and constitutes one of the
main challenges connected with the use of calorimeter-only events.

\subsection{Compton Events}

Although the LAT was not designed as a Compton scattering telescope, it does
in fact have significant acceptance to record Compton interactions---around
$\sim200$~cm$^2$~sr at 5~MeV, with a peak value of $\sim2000$~cm$^2$~sr
around 20~MeV.  
Since the tungsten converters significantly degrade both the spatial and energy
resolution for Compton interactions, extracting useful information out of
them involves selecting the events that are less affected by multiple
scattering---such as front-converting events, or events which convert after
the final tungsten layer and leave signals in the last three tracker layers.
The significant analyzing power of the Compton events for measuring polarization
makes this event class particularly interesting.

\section{New Analysis Techniques}

\passeight will extend the scientific reach of the LAT in areas that are
simply not accessible in the current event-level analysis. The new event
reconstruction and selection are being designed with these scientific targets
in mind and therefore in this section we provide a few illustrative examples.

\subsection{Multi-photon Events}

One of the most striking aspects of the LAT capability compared to prior
missions is its high \emph{shutter speed}. When viewed as a camera,
the LAT has a shutter speed approximately equal to its trigger window width
or $\sim600$~ns and a frame advance time set by the readout dead time or
$\sim26.5$~$\mu$s. When this is considered in combination with its large
effective area the possibility of recording simultaneous photons becomes
tantalizing. It was suggested long before launch that some astrophysical
sources could produce coherent bunches of high-energy gamma rays.
In addition, extraordinary bright, short bright bursts from, for example,
black-hole evaporation could also result in multi-photon events.
However, searches for such exotic events are not possible with the
current reconstruction algorithms. The lack of calorimeter clustering along
with a background rejection tuned on single-photon events almost completely
kills any efficiency the LAT might have to see such events.    
With the re-write of the LAT event-level analysis currently underway, both
of these deficiencies are being addressed. The new calorimeter clustering
algorithm for the first time recognizes and separates distinct energy
depositions within it and this, coupled with the new tracker pattern
recognition, will enable a search for multi-photon events. The Monte Carlo
generator has been re-worked to allow for the generation of such events as well.
This allows testing of the new reconstruction and will guide the design of
analysis cuts to select potential candidates.

\subsection{Polarization measurements}

The idea of using the azimuthal distribution of the electron-positron opening
plane to perform gamma-ray polarimetry in the pair-production regime dates back
more than 60 years \citep{REF.1950.Polarimetry} and has been extensively
studied theoretically. The main limiting factor is that even fully polarized
radiation only gives an overall 10--20\% modulation. In addition to that,
in a typical pair-conversion telescope this modulation is strongly suppressed
(exponentially with the converter thickness) due to multiple Coulomb scattering.

Fermi, thanks to its good hit resolution and large effective area has a much
larger sensitivity to polarimetry than any of its predecessors. The
LAT polarimetric capabilities are, for the first time, being investigated in
detail in the context of \passeight---particularly in conjunction with the
possibility of using gamma-ray conversions in the silicon detectors in order
to limit the effect of the multiple scattering. Preliminary studies indicate
that we might be able to provide meaningful information on the linear
polarization for the strongest gamma-ray sources.

\subsection{Event-by-event Errors\label{subsec:event_by_event_errors}}

The image of sources produced by the LAT is strongly energy-dependent: the
PSF improves roughly as $E^{-0.8}$ in the multiple-scattering regime
(i.e., below $\sim 10$~GeV). There are other factors influencing the image
resolution, namely the conversion point in the tracker (the layout of the
tungsten radiators is such that there is approximately a factor of two
difference in angular resolution between front- and back-converting events)
and the off-axis angle.

The IRFs do capture to some extent all these effects, but they are really
averages of all the various event types which can occur within broad
categories. On the other hand the reconstruction, in addition to providing the
direction of the incoming gamma-rays, also calculates the errors in the form of
an event-by-event covariance matrix deduced from the hit composition and the
material crossed by the initial electron and positron tracks. For example,
if a trajectory is missing an early hit due to the track passing through a gap
between silicon sensors, the associated PSF is degraded by up to a factor of
two. In addition, the current IRF formalism assumes the error for each event is
circular on the sky, an assumption which becomes increasing inaccurate far off
axis where the real photon error is highly elliptical.

In order to exploit the full potential of the LAT in terms of pointing
resolution, \passeight will make the full event-by-event covariant information
available in a form that can be readily used for science analysis.
While an event-by-event analysis of the entire sky is unlikely to be
feasible due to computational limitations, we anticipate that event by event
errors might be key to specific analyses, such as the search for pair-halo
effect in AGNs. More generally, they will be beneficial for source localization
(especially for short transients) and for studying extended sources.

\section{Conclusions\label{sec:conclusions}}

Pass 8 will come close to realizing the full scientific potential of the
\Fermi Large Area Telescope. It incorporates the knowledge gained from the
prime phase of the mission and completes the analysis that was time-limited
prior to launch. The basic ingredients of the new event simulation and
reconstruction are in place and ready to serve as input into the new background
rejection chain which is now being developed. We anticipate \passeight will
be ready to be tested on real science analyses around the end of 2013.

We anticipate that many of the performance improvements will be beneficial to
all science analyses: larger acceptance, better high-energy PSF, lower
backgrounds and better control over the systematic uncertainties.

The new event reconstruction will allow us to extend the energy reach of the
LAT both below 100~MeV (which is of interest for many scientific targets)
and above 1~TeV (diffuse gamma and cosmic-ray electron spectra).
The extended event classes will provide significant enhancements in the
acceptance for specific analyses. Finally, \passeight will allow for new science
topics which are precluded by the current event-level analysis, such as the
search for multi-photon events and $\gamma$-ray polarization measurements.

\bigskip 

\begin{acknowledgments}
  The \Fermi-LAT Collaboration acknowledges generous ongoing support
  from a number of agencies and institutes that have supported both the
  development and the operation of the LAT as well as scientific data analysis.
  These include the National Aeronautics and Space Administration and the
  Department of Energy in the United States, the Commissariat \`a l'Energie
  Atomique and the Centre National de la Recherche Scientifique / Institut
  National de Physique Nucl\'eaire et de Physique des Particules in France,
  the Agenzia Spaziale Italiana and the Istituto Nazionale di Fisica Nucleare
  in Italy, the Ministry of Education, Culture, Sports, Science and Technology
  (MEXT), High Energy Accelerator Research Organization (KEK) and Japan
  Aerospace Exploration Agency (JAXA) in Japan, and the K.~A.~Wallenberg
  Foundation, the Swedish Research Council and the Swedish National Space
  Board in Sweden.

  Additional support for science analysis during the operations phase is
  gratefully acknowledged from the Istituto Nazionale di Astrofisica in Italy
  and the Centre National d'\'Etudes Spatiales in France.
\end{acknowledgments}

\bigskip 

\bibliographystyle{apsrev}
\bibliography{fermisymp2012_pass8overview}

\begin{thebibliography}{7}
\expandafter\ifx\csname natexlab\endcsname\relax\def\natexlab#1{#1}\fi
\expandafter\ifx\csname bibnamefont\endcsname\relax
  \def\bibnamefont#1{#1}\fi
\expandafter\ifx\csname bibfnamefont\endcsname\relax
  \def\bibfnamefont#1{#1}\fi
\expandafter\ifx\csname citenamefont\endcsname\relax
  \def\citenamefont#1{#1}\fi
\expandafter\ifx\csname url\endcsname\relax
  \def\url#1{\texttt{#1}}\fi
\expandafter\ifx\csname urlprefix\endcsname\relax\def\urlprefix{URL }\fi
\providecommand{\bibinfo}[2]{#2}
\providecommand{\eprint}[2][]{\url{#2}}

\bibitem[{\citenamefont{{Ackermann} et~al.}(2012)\citenamefont{{Ackermann},
  {Ajello}, {Albert}, {Allafort}, {Atwood}, {Axelsson}, {Baldini}, {Ballet},
  {Barbiellini}, {Bastieri} et~al.}}]{REF.2012.P7PerfPaper}
\bibinfo{author}{\bibfnamefont{M.}~\bibnamefont{{Ackermann}}},
  \bibinfo{author}{\bibfnamefont{M.}~\bibnamefont{{Ajello}}},
  \bibinfo{author}{\bibfnamefont{A.}~\bibnamefont{{Albert}}},
  \bibinfo{author}{\bibfnamefont{A.}~\bibnamefont{{Allafort}}},
  \bibinfo{author}{\bibfnamefont{W.~B.} \bibnamefont{{Atwood}}},
  \bibinfo{author}{\bibfnamefont{M.}~\bibnamefont{{Axelsson}}},
  \bibinfo{author}{\bibfnamefont{L.}~\bibnamefont{{Baldini}}},
  \bibinfo{author}{\bibfnamefont{J.}~\bibnamefont{{Ballet}}},
  \bibinfo{author}{\bibfnamefont{G.}~\bibnamefont{{Barbiellini}}},
  \bibinfo{author}{\bibfnamefont{D.}~\bibnamefont{{Bastieri}}},
  \bibnamefont{et~al.}, \bibinfo{journal}{\apjs}
  \textbf{\bibinfo{volume}{203}}, \bibinfo{eid}{4} (\bibinfo{year}{2012}),
  \eprint{1206.1896}.

\bibitem[{\citenamefont{{Atwood} et~al.}(2009)\citenamefont{{Atwood}, {Abdo},
  {Ackermann}, {Althouse}, {Anderson}, {Axelsson}, {Baldini}, {Ballet}, {Band},
  {Barbiellini} et~al.}}]{REF.2009.LATPaper}
\bibinfo{author}{\bibfnamefont{W.~B.} \bibnamefont{{Atwood}}},
  \bibinfo{author}{\bibfnamefont{A.~A.} \bibnamefont{{Abdo}}},
  \bibinfo{author}{\bibfnamefont{M.}~\bibnamefont{{Ackermann}}},
  \bibinfo{author}{\bibfnamefont{W.}~\bibnamefont{{Althouse}}},
  \bibinfo{author}{\bibfnamefont{B.}~\bibnamefont{{Anderson}}},
  \bibinfo{author}{\bibfnamefont{M.}~\bibnamefont{{Axelsson}}},
  \bibinfo{author}{\bibfnamefont{L.}~\bibnamefont{{Baldini}}},
  \bibinfo{author}{\bibfnamefont{J.}~\bibnamefont{{Ballet}}},
  \bibinfo{author}{\bibfnamefont{D.~L.} \bibnamefont{{Band}}},
  \bibinfo{author}{\bibfnamefont{G.}~\bibnamefont{{Barbiellini}}},
  \bibnamefont{et~al.}, \bibinfo{journal}{\apj} \textbf{\bibinfo{volume}{697}},
  \bibinfo{pages}{1071} (\bibinfo{year}{2009}), \eprint{0902.1089}.

\bibitem[{\citenamefont{{Rochester} et~al.}(2010)\citenamefont{{Rochester},
  {Usher}, {Johnson}, and {Atwood}}}]{REF.2010.Pass8}
\bibinfo{author}{\bibfnamefont{L.}~\bibnamefont{{Rochester}}},
  \bibinfo{author}{\bibfnamefont{T.}~\bibnamefont{{Usher}}},
  \bibinfo{author}{\bibfnamefont{R.~P.} \bibnamefont{{Johnson}}},
  \bibnamefont{and} \bibinfo{author}{\bibfnamefont{B.}~\bibnamefont{{Atwood}}},
  \bibinfo{journal}{ArXiv e-prints}  (\bibinfo{year}{2010}),
  \eprint{1001.5005}.

\bibitem[{\citenamefont{{Bruel}}(2012)}]{REF.2012.EnergyRecon}
\bibinfo{author}{\bibfnamefont{P.}~\bibnamefont{{Bruel}}},
  \bibinfo{journal}{Journal of Physics Conference Series}
  \textbf{\bibinfo{volume}{404}}, \bibinfo{pages}{012033}
  (\bibinfo{year}{2012}), \eprint{1210.2558}.

\bibitem[{\citenamefont{Hoecker et~al.}(2007)\citenamefont{Hoecker, Speckmayer,
  Stelzer, Therhaag, von Toerne, and Voss}}]{REF.2007.TMVA}
\bibinfo{author}{\bibfnamefont{A.}~\bibnamefont{Hoecker}},
  \bibinfo{author}{\bibfnamefont{P.}~\bibnamefont{Speckmayer}},
  \bibinfo{author}{\bibfnamefont{J.}~\bibnamefont{Stelzer}},
  \bibinfo{author}{\bibfnamefont{J.}~\bibnamefont{Therhaag}},
  \bibinfo{author}{\bibfnamefont{E.}~\bibnamefont{von Toerne}},
  \bibnamefont{and} \bibinfo{author}{\bibfnamefont{H.}~\bibnamefont{Voss}},
  \bibinfo{journal}{PoS} \textbf{\bibinfo{volume}{ACAT}}, \bibinfo{pages}{040}
  (\bibinfo{year}{2007}), \eprint{physics/0703039}.

\bibitem[{\citenamefont{{Pelassa} et~al.}(2010)\citenamefont{{Pelassa},
  {Preece}, {Piron}, {Omodei}, {Guiriec}, {Fermi LAT}, and {GBM
  collaborations}}}]{REF.2010.LLE}
\bibinfo{author}{\bibfnamefont{V.}~\bibnamefont{{Pelassa}}},
  \bibinfo{author}{\bibfnamefont{R.}~\bibnamefont{{Preece}}},
  \bibinfo{author}{\bibfnamefont{F.}~\bibnamefont{{Piron}}},
  \bibinfo{author}{\bibfnamefont{N.}~\bibnamefont{{Omodei}}},
  \bibinfo{author}{\bibfnamefont{S.}~\bibnamefont{{Guiriec}}},
  \bibinfo{author}{\bibfnamefont{f.~t.} \bibnamefont{{Fermi LAT}}},
  \bibnamefont{and} \bibinfo{author}{\bibnamefont{{GBM collaborations}}},
  \bibinfo{journal}{ArXiv e-prints}  (\bibinfo{year}{2010}),
  \eprint{1002.2617}.

\bibitem[{\citenamefont{{Yang}}(1950)}]{REF.1950.Polarimetry}
\bibinfo{author}{\bibfnamefont{C.~N.} \bibnamefont{{Yang}}},
  \bibinfo{journal}{Physical Review} \textbf{\bibinfo{volume}{77}},
  \bibinfo{pages}{722} (\bibinfo{year}{1950}).

\end{thebibliography}

\end{document}